\documentclass[conference, 10pt]{IEEEtran}

\usepackage{cite}
\usepackage{amsmath,amssymb,amsfonts}
\usepackage{algorithmic}
\usepackage{graphicx}
\usepackage{textcomp}
\usepackage{xcolor}
\def\BibTeX{{\rm B\kern-.05em{\sc i\kern-.025em b}\kern-.08em
    T\kern-.1667em\lower.7ex\hbox{E}\kern-.125emX}}

\usepackage[cmintegrals]{newtxmath}
\usepackage{bm} %need to manually load newtxmath before bm
\usepackage{enumitem}
\usepackage{soul}
\setstcolor{purple}
\usepackage{microtype}
\usepackage{algorithm2e}
\usepackage{balance}

\graphicspath{{.}{./images/}}

\newtheorem{theorem}{Theorem}
\newtheorem{lemma}{Lemma}

\newtheorem{remark}{Remark}

\usepackage{etoolbox}
\makeatletter
\patchcmd{\@begintheorem}{\textit}{\textbf}{}{}
\patchcmd{\@begintheorem}{\itshape}{\bfseries}{}{}
\makeatother

\begin{document}
\title{Network Integrated Sensing and Communication} 

\author{
  \IEEEauthorblockN{Edward Andrews$^\ast$, Lawrence Ong$^\ast$, Duy T.\ Ngo$^\ast$, Yao Liu$^\dagger$, Min Li$^\ddagger$}
  \thanks{Lawrence Ong is the corresponding author.}
\IEEEauthorblockA{$^\ast$The University of Newcastle, Callaghan NSW 2308, Australia.\\ Emails: edward.andrews@uon.edu.au,\{lawrence.ong, duy.ngo\}@newcastle.edu.au\\
$^\dagger$Hangzhou Dianzi University, Hangzhou 310018, China, Email: yao.liu@hdu.edu.cn\\
$^\ddagger$Zhejiang University, Hangzhou 310027, China, Email: min.li@zju.edu.cn}
}
% \IEEEauthorblockN{Edward Andrews, Lawrence Ong, Duy T. Ngo}\thanks{Lawrence Ong is the corresponding author.}
% \IEEEauthorblockA{The University of Newcastle, Callaghan NSW 2308, Australia\\
% Email: Edward.Andrews@uon.edu.au;\\\{lawrence.ong, duy.ngo\}@newcastle.edu.au}
% \and
% \IEEEauthorblockN{Yao Liu, Min Li}
% \IEEEauthorblockA{Zhejiang University, Hangzhou 310027, China\\
% Email: \{yao.liu, min.li\}@zju.edu.cn}
% }

% %%% Several authors with up to three affiliations:
% \author{%
%   \IEEEauthorblockN{Author 1}\thanks{This work is supported by XXX.}
%   \IEEEauthorblockA{University\\
%     Email: }
%   \and
%   \IEEEauthorblockN{Author 2}
%   \IEEEauthorblockA{University\\
%     Email: }
%       \and
%   \IEEEauthorblockN{Author 3}
%   \IEEEauthorblockA{University\\
%     Email: }
%           \and
%   \IEEEauthorblockN{Author 4}
%   \IEEEauthorblockA{University\\
%     Email: }
    
% }

\IEEEoverridecommandlockouts
\IEEEaftertitletext{\vspace{-1\baselineskip}}

\maketitle

%%%%%%
%% Abstract: 
%% If your paper is eligible for the student paper award, please add
%% the comment "THIS PAPER IS ELIGIBLE FOR THE STUDENT PAPER
%% AWARD." as a first line in the abstract. 
%% For the final version of the accepted paper, please do not forget
%% to remove this comment!
%%
\begin{abstract} 
Integrated sensing and communication (ISAC) is a cornerstone technology for 6G networks, offering unified support for high-rate communication and high-accuracy sensing. While existing literature extensively covers link-level designs, the transition toward large-scale deployment necessitates a fundamental understanding of network-level performance. This paper investigates a network ISAC model where a source node communicates with a destination via a relay network, while intermediate nodes concurrently perform cooperative sensing over specific spatial regions. We formulate a novel optimization framework that captures the interplay between multi-node routing and sensing coverage. For a one-dimensional path network, we provide an analytical characterization of the complete sensing--throughput region. Extending this to general network topologies, we establish that the sensing--throughput Pareto boundary is piecewise linear and provide physical interpretations for each segment. Our results reveal the fundamental trade-offs between sensing coverage and communication routing, offering key insights for the design of future 6G heterogeneous networks.
\end{abstract}


\section{Introduction} \label{sec:intro}
Integrated sensing and communication (ISAC) has been identified as one of the core technologies for the sixth generation (6G) of cellular networks, supporting both high-rate communication and high-accuracy sensing to enable a wide range of intelligent applications~\cite{LIU2022,ZHANG2022}. By employing unified waveforms and jointly utilizing hardware and spectrum resources, ISAC can substantially reduce hardware and energy costs while improving spectral efficiency~\cite{Meng2024,Lu2024}.

%\par 

Extensive studies have been conducted on the key designs for ISAC, including waveform design~\cite{liyanaarachchi2021optimized,rou2024orthogonal}, beamforming optimization~\cite{he2022energy,hua2023optimal}, and signal processing~\cite{wu2022integrating,wei2023iterative}, as well as experimental validations~\cite{wang2019first,ma2021spatial,temiz2023experimental}. Moreover, ISAC has been combined with other emerging technologies, such as reconfigurable intelligent surfaces (RIS)~\cite{luo2022joint} and unmanned aerial vehicle (UAV)~\cite{meng2023uav}, further demonstrating the benefits of unified system design over conventional separated approaches.

To guide the design and optimization of ISAC systems, a growing body of information-theoretic research has focused on the fundamental tradeoffs between communication and sensing performance. These studies have characterized optimal or achievable tradeoff regions for various network topologies, such as point-to-point channels~\cite{Kobayashi2018,liu2022information}, multiple access channels~\cite{kobayashi2019joint,ahmadipour2023information,liu2025fundamental}, broadcast channels~\cite{Ahmadipour-TIT-2024}, and interference channels~\cite{liu2025JSAIT}. Moreover, the performance limits of ISAC under practical system constraints, such as multiple antennas~\cite{xiong2023fundamental} and information security~\cite{gunlu2023secure}, have also been investigated. Despite these advances, most existing studies primarily focus on node-level or link-level performance analysis and optimization.

As ISAC moves toward large-scale deployment in 6G heterogeneous networks, it becomes necessary to go beyond link-level analysis and investigate network-level performance. Several preliminary studies have begun to explore this perspective. In particular,~\cite{meng2024network} and \cite{meng2025network} have focused on monostatic sensing and employed stochastic geometry to characterize the joint sensing-and-communication performance under base-station coordination or cooperation. Corresponding optimization problems are formulated to tune key network-level parameters, aiming to maximize the overall network performance with respect to joint sensing-and-communication metrics. Furthermore, \cite{qiu-isit-2025} has investigated the relationship between sensing regions and communication performance in an ad-hoc network with dual-functional nodes. However, their definition of the sensing region is restricted to a sensing distance around each node rather than a specific spatial area, and their results are mainly presented in terms of scaling laws.

In this paper, we focus on network ISAC, where a source node delivers messages to a destination through a relay network, with some intermediate nodes in the network simultaneously performing cooperative sensing (characterized by the sensing region) while relaying communication messages. Our objective is to characterize the fundamental tradeoff between communication throughput and sensing performance for this network. The main research contributions are summarized as follows:
\begin{enumerate}
    \item We formulate a new network-level ISAC problem that captures (i) multiple-node network operation, (ii) communications via routing messages over relay nodes, (iii) sensing regions induced by links therein, and (iv) trade-off between sensing and communication in each link.
    \item For the one-dimensional path ISAC network, we analytically characterize the complete sensing--throughput region.
    \item For the general ISAC network, we characterize (i) the shape of the sensing--throughput region, which is piecewise linear, including the physical interpretation of each linear segment on the Pareto boundary, and (ii) the fundamental interplay between sensing and routing for communication.
\end{enumerate}

\section{Modeling and Problem Formulation of Network ISAC}\label{sec:system-model}
We use the following notation: 
$\mathbb{Z}^+$ denotes the set of natural numbers, 
$[a:b] := \{a, a+1, \dotsc, b\}$ for $a,b\in\mathbb{Z}^+$ such that $a < b$,
and
$\mathbb{R}^+_0$ denotes the set of non-negative real numbers.

\subsection{Network Model}
An ISAC network is defined by a directed graph $G= (V,E)$ where:
\begin{itemize}
    \item $V = [1:N]$ is the set of nodes (vertices).
    \item $E \subseteq V \times V$ is the set of directed links (edges), representing ISAC links that are capable of communicating data and sensing the surrounding environment.
    \item  Links occur in both directions, i.e., if $(i,j) \in E$, then $(j,i) \in E$. We use the undirected link $\{i,j\}$ to denote the pair of links between nodes $i$ and $j$ in both directions. Let $U :=\{\{i,j\}: (i,j) \in E\}$ denote the set of undirected links. 
    \item $(i,i) \notin E$. So, by definition, $|E| = 2 |U|$.
    \item $T_x \in V$ denotes the communications source node.
    \item $R_x \in V$ denotes the communications sink node.
\end{itemize} 

Each undirected ISAC link has a fixed \textit{capacity}---this term is to be treated in a general sense that encompasses communication and sensing. Denote the capacity of the undirected link $\{i,j\} \in U$ by $c_{\{i,j\}} \in \mathbb{R}^+_0$ (average bits per second). The capacity of the undirected link $\{i,j\}$ is time-shared between the links in both directions, i.e.,
\begin{equation}\label{eq:link-capacity-splitting}
    c_{i,j} + c_{j,i} = c_{\{i,j\}}, \forall \{i,j\} \in U,
\end{equation} where $c_{i,j} \in \mathbb{R}^+_0$ is the capacity of the directed link $(i,j)$.

As the system is intended to be analyzed from a network-level perspective, considerations of wireless signal processing (e.g., medium access control, interference, fading, waveform design) are abstracted to lower layers of the Internet protocol stack. This gives rise to the above-mentioned independent links with fixed capacity.

\subsection{Link Model}
For point-to-point AWGN monostatic\footnote{In monostatic sensing, the transmitter senses the environment from echoed signals.} ISAC, the optimal trade-off between sensing and communication has been characterised~\cite{Ahmadipour-TIT-2024}. Specifically, the sensing fidelity is taken to be the expected average per-block distortion between the state sequence and the estimated state sequence. The optimal trade-off is non-linear.

In this work, we consider a linear trade-off between sensing and communication for each ISAC link, i.e.,
\begin{equation}\label{eq:trade-off}
    f_{i,j}+s_{i,j}\leq c_{i,j}, \forall (i,j) \in E,
\end{equation}
where
\begin{itemize}
    \item $f_{i,j} \in \mathbb{R}^+_0$ (average bits per second) is the rate of communication over the directed link $(i,j)$. By definition, $f_{i,T_x} = 0,\forall (i,T_x) \in E$ and $f_{R_x,j} = 0,\forall (R_x,j) \in E$.
    \item $s_{i,j} \in \mathbb{R}^+_0$ (average bits per second) is the sensing rate obtained over the directed link $(i,j)$.
\end{itemize}

Eqn. \eqref{eq:trade-off} models an ISAC link that is time-shared between communication and sensing. While time-sharing between communication and sensing is suboptimal~\cite{Ahmadipour-TIT-2024} for the point-to-point AWGN monostatic sensing, such a linear trade-off function at the link level allows the generation of interpretable and instructive results. The network ISAC framework proposed in this paper forms a basis on which other non-linear trade-off link models for sensing and communication can be readily substituted.%\footnote{\color{red}LO: How about this now? We do not say anything about the result.}
%The simplification of this assumed link trade-off approach forms a basis for further analyses, where other point-to-point ISAC models can be substituted for the network-level results obtained here \footnote{DN: not very clear to me. Do we mean ``other point-to-point ISAC models can be used instead of the linear model and we still can obtain similar network-level results as in this paper?".}. 

\subsection{Network Sensing Fidelity}

While the sensing fidelity for point-to-point ISAC has been characterized using expected distortion~\cite{Ahmadipour-TIT-2024} or the estimation rate\footnote{Estimation rate is defined as the mutual information between the state to be estimated and the observation}~\cite{bell-tit-1993}, network ISAC requires an area-based sensing metric. Recently, Qiu et al.~\cite{qiu-isit-2025} defined the sensing fidelity as the minimum sensing distance among all the nodes in network ISAC, where the sensing distance of a node is a function of its transmitted power.

In this paper, since the ISAC links have been abstracted from the lower layers, each link captures sensing in its proximity. A collection of the links then gives rise to a sensing region within which the links reside. With that, we model the sensing fidelity within a region of interest as the sum of the sensing rates of all the links within that region. We note that the sensing fidelity is application-dependent, and the metric we adopted may not always be applicable in general. For instance, if several links are tracking the same target (e.g., sensing of ambient temperature) and hence the sensing observations are correlated, the overall sensing fidelity is unlikely to be the sum of individual measures. Rather, our adopted metric applies to independent observations, e.g., tracking the number of cars on different roads or monitoring wildlife activities in a region.

Formally, let us define $A \subseteq V$ as the set of nodes that map out the area of interest for sensing. Then, $U(A):= \{\{i,j\} \in U: i,j \in A\}$ captures the set of undirected links that collectively sense the region. Similarly, we define $E(A):= \{(i,j)\in E: \forall i,j\in A\}$ as the set of directed links that capture the sensing region. The sensing fidelity is then defined as 
\begin{equation}\label{eq:sensing}
    s := \sum_{u \in U(A)} s_u := \sum_{\{i,j\} \in U(A)} (s_{i,j} + s_{j,i}) = \sum\limits_{e \in E(A)} s_e.
\end{equation}

\subsection{Communication Throughput}

The communication throughput, $F$, is defined as the rate of communication from the source node to the sink node through the network. A set of communication-rate assignments $\{f_{i,j}: \forall (i,j) \in E\}$ for all the links must satisfy the conservation of message flows\footnote{This view of message flows is more restrictive than the general network-coding formulation, where the outgoing links are functions of the incoming links. However, for single-source single-sink networks, routing (i.e., treating information as flows or bit pipes) is optimal.}, i.e.,
\begin{equation}\label{eq:flow}
    \sum\limits_{i:(i,j)\in E}f_{i,j} - \sum\limits_{k:(j,k)\in E}f_{j,k} = 0, \forall j \in V \setminus \{T_x,R_x\}.
\end{equation}
With that, the throughput is given by
\begin{equation}\label{eq:throughput}
     f = \sum_{j:(T_x,j)\in E}f_{T_x,j}.
\end{equation}
For a single-source single-sink network, the maximum throughput, $f^*$, is given by the max-flow min-cut theorem.

\subsection{Problem Formulation}

Consider an ISAC network $G = (V,E)$, undirected link capacities $\{c_u: u \in U\}$, and a sensing area of interest $A \subseteq V$.

A set of sensing-rate and communication-rate assignments $\{s_e,f_e \in \mathbb{R}^+_0: e\in E\}$ are said to be \textit{valid} if they satisfy link-spitting constraint~\eqref{eq:link-capacity-splitting} and linear trade-off~\eqref{eq:trade-off} according to some link-rate splitting assignments $\{c_e \in \mathbb{R}^+_0: e\in E\}$, as well as the conservation of flows \eqref{eq:flow}.

A sensing--throughput pair $(s, f)$ is said to be \textit{feasible} if it is \textit{achieved} by a set of valid sensing-rate and communication-rate assignments according to \eqref{eq:sensing} and \eqref{eq:throughput}.

\textbf{Problem}:  To find the set of all feasible sensing--throughput pairs, which we call the \textit{sensing--throughput region}, $R \subseteq \mathbb{R}^+_0 \times \mathbb{R}^+_0$.

\begin{remark}
    We assume that the amount of sensed data is insignificant compared to the communications data. Sensing data, such as temperature measurements and object counts and positions, are generally characterized by low data rates. Accordingly, the transmission of sensing data acquired via the ISAC links to a sensing fusion centre for further processing is neglected.
\end{remark}

\section{Basic Properties}

This section presents several simple yet useful results that will be used throughout the paper.

\begin{lemma} \label{lemma:R-convex}
$R$ is convex.
\end{lemma}
\begin{IEEEproof}
    Let $(s^{(\alpha)},f^{(\alpha)}) \in R$, achievable using link-splitting assignments $\{c^{(\alpha)}_e: e\in E\}$, communication-rate assignments $\{f^{(\alpha)}_e: \forall e\in E\}$, and sensing-rate assignments $\{s^{(\alpha)}_e: \forall e\in E\}$, satisfy \eqref{eq:link-capacity-splitting}, \eqref{eq:trade-off} and \eqref{eq:flow}. Also, let $(s^{(1-\alpha)},f^{(1-\alpha)}) \in R$, achievable using link-splitting assignments $\{c^{(1-\alpha)}_e: e\in E\}$, communication-rate assignments $\{f^{(1-\alpha)}_e: \forall e\in E\}$, and sensing-rate assignments $\{s^{(1-\alpha)}_e: \forall e\in E\}$, satisfy \eqref{eq:link-capacity-splitting}, \eqref{eq:trade-off} and \eqref{eq:flow}.

    We will show that, using time sharing, $(\alpha s^{(\alpha)} + (1-\alpha)s^{(1-\alpha)}, \alpha f^{(\alpha)} + (1-\alpha)f^{(1-\alpha)}) \in R$, for any $\alpha \in [0,1]$, and hence $R$ is convex.

    Let the new communication-rate assignments be $f_e = \alpha f^{(\alpha)}_e + (1-\alpha)f^{(1-\alpha)}_e$, $\forall (i,j) \in E$. Then, for all $j \in V \setminus \{T_x,R_x\}$,
    \begin{subequations}
    \begin{align}
        & \sum\limits_{i:(i,j)\in E}f_{i,j} - \sum\limits_{k:(j,k)\in E}f_{j,k} \nonumber\\
        &= \alpha \left( \sum\limits_{i:(i,j)\in E}f^{(\alpha)}_{i,j} - \sum\limits_{k:(j,k)\in E}f^{(\alpha)}_{j,k} \right)\nonumber\\
        &\quad + (1-\alpha) \left( \sum\limits_{i:(i,j)\in E}f^{(1-\alpha)}_{i,j} - \sum\limits_{k:(j,k)\in E}f^{(1-\alpha)}_{j,k} \right)\\
        &=0,
    \end{align}
    \end{subequations}
    where the last equality holds as the rate assignments $\{f^{(\alpha)}_e: e \in E\}$ and $\{f^{(1-\alpha)}_e: e \in E\}$ both satisfy \eqref{eq:flow}.

    Also, let the new link-splitting assignments be $c_e = \alpha c^{(\alpha)}_e + (1-\alpha) c_e^{(1-\alpha)}, \forall e\in E$. So, for each $\{i,j\} \in U$,
    \begin{subequations}
    \begin{align}
        &c_{i,j} + c_{j,i} \nonumber\\
        &= \alpha\left( c^{(\alpha)}_{i,j} + c^{(\alpha)}_{j,i} \right) + (1-\alpha)\left( c^{(1-\alpha)}_{i,j} + c^{(1-\alpha)}_{j,i} \right)\\
        &= \alpha c_{\{i,j\}} + (1-\alpha) c_{\{i,j\}} \label{eq:a}\\
        &= c_{\{i,j\}},
    \end{align}
    \end{subequations}
    where \eqref{eq:a} holds as the splitting assignments $\{c^{(\alpha)}_e: e \in E\}$ and $\{c^{(1-\alpha)}_e: e\in E\}$ both satisfy \eqref{eq:link-capacity-splitting}.
    
    Finally, let the new sensing-rate assignments be $s_e = \alpha s^{(\alpha)}_e + (1-\alpha)s^{(1-\alpha)}_e$, $\forall (i,j) \in E$. Then, for each $e \in E$,
    \begin{subequations}
    \begin{align}
        &f_e + s_e \nonumber \\
        &= \alpha \left( f^{(\alpha)}_e + s^{(\alpha)}_e \right) + (1-\alpha) \left( f^{(1-\alpha)}_e + s^{(1-\alpha)}_e \right)\\
        &\leq \alpha c^{(\alpha)}_e + (1-\alpha)c^{(1-\alpha)}_e \label{eq:b}\\
        &= c_e,
    \end{align}
    \end{subequations}
    where \eqref{eq:b} is obtained as the rate assignments $\{s^{(\alpha)}_e,f^{(\alpha)}_e: e \in E\}$ and $\{s^{(1-\alpha)}_e,f^{(1-\alpha)}_e: e \in E\}$ both satisfy \eqref{eq:trade-off}.

    Since the new link and rate assignments are valid, we have $(s,f) \in R$, where $f = \sum_{j:(T_x,j)\in E}f_{T_x,j} = \sum_{j:(T_x,j)\in E}\left( \alpha f^{(\alpha)}_{T_x,j} + (1-\alpha) f^{(1-\alpha)}_{T_x,j}\right) = \alpha f^{(\alpha)} + (1-\alpha)f^{(1-\alpha)}$ and $s = \sum\limits_{e \in E(A)} s_e = \sum\limits_{e \in E(A)} \left( \alpha s^{(\alpha)}_e + (1-\alpha) s^{(1-\alpha)}_e \right) = \alpha s^{(\alpha)} + (1-\alpha)s^{(1-\alpha)}$. This completes the proof.
\end{IEEEproof}

\begin{lemma}\label{lemma:s-to-0}
    If $(s,f) \in R$, then $(0,f) \in R$.
\end{lemma}
\begin{IEEEproof}
    Suppose that $(s,f)$ is achieved by valid set of communication-rate and sensing-rate assignments. Setting $s_e=0, \forall e \in E(A)$ maintains the validity of the assignments, and therefore achieves $(0,f)$.
\end{IEEEproof}

\begin{lemma}\label{lemma:f-to-0}
    If $(s,f) \in R$, then $(s,0) \in R$.
\end{lemma}
\begin{IEEEproof}
    The proof is similar to that of Lemma~\ref{lemma:s-to-0} by setting $f_e = 0, \forall e \in E$ and checking that doing so does not violate the validity of the rate assignments.
\end{IEEEproof}

\begin{lemma}\label{lemma:smaller-rate}
    If $(s,f) \in R$, then $(s - \sigma, f - \delta) \in R$ for any $\sigma \in (0,s]$ and $\delta \in (0,f]$.
\end{lemma}

\begin{IEEEproof}
    This follows from Lemmas~\ref{lemma:R-convex}, \ref{lemma:s-to-0}, and \ref{lemma:f-to-0}.
\end{IEEEproof}
    % Suppose that $(f,s) \in R$, then there must be a set of sensing-rate assignments $\{s_e: e \in E\}$ that support the sensing fidelity $s$ according to \eqref{eq:sensing} and a valid set of communication-rate assignments $\{f_e: e \in E\}$ that support the throughput $f$ according to \eqref{eq:throughput}. Also, all assignments meet constraint~\eqref{eq:trade-off}.
    
    % First, according to \eqref{eq:sensing}, the sensing-rate allocations can always be reduced by a total amount of $\sigma \in (0,s]$, while maintaining each sensing-rate allocation to be non-negative. And doing so will not change constraint~\eqref{eq:trade-off}. This proves $(f, s - \sigma) \in R$.

    % Now, let $c_\text{min} := \min \{f_{i,j}: f_{i,j}>0, (i,j) \in E\}$. Note that $c_\text{min} \leq f$. Due to the conservation of flow~\eqref{eq:flow}, we can always find a path from the source node to the destination node, denoted by $P = \{(T_x, i_1),(i_1,i_2), (i_2, i_3), \dotsc, (i_{k-1},i_k), (i_k,R_x)\}$, where $T_x \neq i_1 \neq i_2 \neq \dotsm \neq i_k \neq R_x$, that sasitfies $f_e \geq c_\text{min}, \forall e \in P$. In fact, all the paths ``used'' for communication has this property. Consequently, we can reduce the throughput by $\delta' \in (0, c_\text{min}]$ by reducing the communication rate in each link in $P$ by $\delta'$. By construction, the conservation of flow~\eqref{eq:flow} and constraint~\eqref{eq:trade-off} remain satisfied. Repeating this argument until $f = c_\text{min}$, we can reduce the throughput by $\delta \in (0,f]$, and therefore $(f-\delta, s) \in R$.

Denoting the maximum throughput by $f^*$ and the maximum sensing fidelity by $s^*$, we have the following results:

\begin{lemma}\label{lemma:max-throughput}
    $(0,f^*) \in R$, i.e., the maximum throughput is attained without sensing. 
\end{lemma}
\begin{IEEEproof}
    This follows from Lemma~\ref{lemma:s-to-0}.
\end{IEEEproof}

\begin{lemma}\label{lemma:max-sensing}
    $(s^*,0) \in R$, i.e., the maximum sensing fidelity is attained without communication.
\end{lemma}

\begin{IEEEproof}
    This follows from Lemma~\ref{lemma:f-to-0}.
\end{IEEEproof}

\begin{lemma}\label{lemma:max-sensing-value}
    $s^* = \sum_{u \in U(A)} c_u$.
\end{lemma}
\begin{IEEEproof}
    Allocating all link capacities for sensing $s_e = c_e, \forall e \in E(A),$ maximizes \eqref{eq:sensing}. The allocation of $c_{i,j}$ and $c_{j,i}$ within an undirected link $\{i,j\}$ does not matter.
\end{IEEEproof}

Lemma~\ref{lemma:max-sensing-value} is consistent with Lemma~\ref{lemma:max-sensing} by setting $f_e = 0, \forall e \in E$, i.e., allocating all link capacities for sensing and not communicating. Note that not communicating is sufficient but may not be necessary to maximize sensing.

Consider the following linear program $\mathcal{P}_1(G,\{c_u:u \in U\}, A, T_S)$ for given network $G = (V,E)$, link capacities $\{c_u: u \in U\}$, and sensing region $A \subseteq V$, and a sensing-fidelity target $T_S \in [0,s^*]$:
\begin{subequations}\label{eq:problem_definition}
\begin{align}
\max_{f_e, s_e: e \in E} \,\, & f\\
%\end{align}
\textrm{s.t.} \,\, & \eqref{eq:link-capacity-splitting}, \eqref{eq:trade-off}, \eqref{eq:flow},\\
%\begin{align}
& f_e, s_e \geq 0, \forall e \in E,\label{eq:non-negative-rates} \\
&\sum\limits_{\{i,j\} \in U(A)} s_{\{i,j\}} = T_S. \label{eq:SensingReq}
\end{align}
\end{subequations}

\begin{lemma}\label{lemma:lp}
    The linear program $\mathcal{P}_1(G,\{c_u:u \in U\}, A, T_S)$ finds $\max \{f: (T_S,f) \in R\}$ for a fixed $T_S \in [0:s^*]$f%\footnote{DN: check to see if the notation is correct. \color{red} LO: This looks better?}
    , i.e., the maximum throughput for a given sensing-fidelity target $T_S$.
\end{lemma}
\begin{IEEEproof}
    Each set of sensing-rate and communication-rate assignments $\{s_e,f_e: e \in E\}$ in the feasible set of $\mathcal{P}_1(G,\{c_u:u \in U\}, A, T_S)$ are valid as they satisfy the requirements \eqref{eq:link-capacity-splitting}, \eqref{eq:trade-off}, and \eqref{eq:flow} according to some link-rate splitting assignments $\{c_e:e \in E\}$. Hence, any set of such assignments achieve a feasible throughput--sensing pair  $(T_S,f) \in R$, where $T_S$ is the sensing fidelity fixed by \eqref{eq:SensingReq}.

    Conversely, any feasible throughput--sensing pair $(T_S,f) \in R$ must be achieved by a valid set of communication-rate and sensing-rate assignments, which satisfy the requirements \eqref{eq:link-capacity-splitting}, \eqref{eq:trade-off}, and \eqref{eq:flow} according to some link-rate splitting assignments $\{c_e:e \in E\}$. Hence, the set are a feasible solution to $\mathcal{P}_1(G,\{c_u:u \in U\}, A, T_S)$.

    Consequently, the optimal value of $\mathcal{P}_1(G,\{c_u:u \in U\}, A, T_S)$ must be the largest $f$ for which $(T_S,f) \in R$.
\end{IEEEproof}

\begin{remark}
Due to Lemma~\ref{lemma:lp}, the throughput--sensing region $R$ can be numerically generated by solving the linear program $\mathcal{P}_1(G,\{c_u:u \in U\}, A, T_S)$ for $T_S \in [0,s^*]$.    
\end{remark}

\begin{lemma}\label{lemma:one-link-off}
    $R$ can be obtained by considering only one direction in every undirected link $\{i,j\} \in U$, i.e., either $f_{i,j} = s_{i,j} = 0$ or $f_{j,i} = s_{j,i} = 0$.
\end{lemma}

\begin{IEEEproof}
    Suppose that $(s,f) \in R$, achieved by a valid set of sensing-rate and communication-rate assignments $\{s_e, f_e: e \in E\}$, i.e., they satisfy \eqref{eq:link-capacity-splitting}, \eqref{eq:trade-off}, and \eqref{eq:flow}.

    Now, combining \eqref{eq:link-capacity-splitting} and \eqref{eq:trade-off} gives
    \begin{equation}\label{eq:combined}
        f_{i,j} + f_{j,i} + s_{i,j} + s_{j,i} \leq c_{\{i,j\}}, \forall \{i,j\} \in U.
    \end{equation}

    Consider any undirected link $\{i,j\} \in U$. Pick any $e_1 \in \arg\!\max_{e \in \{ (i,j), (j,i) \}} f_e$, and let $e_2$ be the other link, i.e., $\{ (i,j), (j,i) \} \setminus \{e_1\}$. Let a new set of rate assignments be $\{s'_e,f'_e: e \in E\}$, where for each link $\{i,j\} \in U$,
    \begin{align}
        f'_{e_1} &= \max_{e \in \{ (i,j), (j,i) \}} f_e - \min_{e \in \{ (i,j), (j,i) \}} f_e \label{eq:aa}\\
        f'_{e_2} & = 0 \label{eq:bb}\\
        s'_{e_1} &= s_{i,j} + s_{j,i} \label{eq:cc}\\
        s'_{e_2} & = 0. \label{eq:dd}
    \end{align}

    We see that if $\{s_e, f_e: e \in E\}$ satisfy \eqref{eq:combined} and \eqref{eq:flow}, so do $\{s'_e,f'_e: e \in E\}$. Let $(s',f')$ be the sensing--throughput pair achieved by $\{s'_e,f'_e: e \in E\}$. %\footnote{DN: check this. \color{red}LO: OK now?}
    We see that modifications \eqref{eq:cc}--\eqref{eq:dd} do not change the sensing fidelity. So, $s' = s$. Also, due to the definition that $f_{i,T_x} = 0,\forall (i,T_x) \in E$, we have $f' = \sum_{j:(T_x,j)\in E}f'_{T_x,j} = \sum_{j:(T_x,j)\in E}f_{T_x,j} = f$.
\end{IEEEproof}

\begin{remark}\label{remark:one-link-off}
    As a result of Lemma~\ref{lemma:one-link-off}, for the remainder of the paper, without loss of optimality%\footnote{\color{red}LO: I change WLOG to without loss of optimality, as this restriction is not as general, but we can get the full $R$. Sounds OK?}
    , we assume that, for each undirected link, one of the two directed links has zero communication and sensing rates. Consequently, the entire capacity of each undirected link is allocated to the other directed link, which has non-zero communication and sensing rates.
\end{remark}

\section{The One-Dimensional ISAC Network}

\subsection{Model}
\begin{figure}[t]
    \centering
    \includegraphics[scale=0.7]{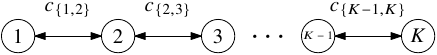}
    \caption{A one-dimensional ISAC network}
    \label{fig:1-DNetworkDiagram}
\end{figure}

We first consider a simplified case of the ISAC network, where we have a one-dimensional path network, i.e., there is only one communications path from source to sink, which must pass through all nodes within the system, as shown in Figure~\ref{fig:1-DNetworkDiagram}. Formally, the nodes are $V=\{1,2,\dots,K\}$, and the links are $E = \{\{i,i+1\}: i \in [1:K-1]\}$. The source node is $T_X = 1$, and the destination node is $R_X = K$. We consider a non-zero sensing area $A \neq \emptyset$ and hence $|U(A)| \geq 1$; otherwise, the ISAC network degenerates to a pure communication network.

 %The Analysis of this system provides insights into the network level trade-off and introduces key concepts that arise from network wide optimisation. Most crucially, one-dimensional networks intuitively introduce the concept of free sensing within a network, where sensing is obtained without detriment to the communications flow, and demonstrates the conditions required for free communications, where the communications flow is achieved without detriment to the sensing data. This system also allows for the development of a closed form solution to the generalised optimisation problem, which provides analytical insights into the optimal sensing-communications boundary for generalised systems.

\subsection{Results}

First, we show that all ``forward'' links carry the same communication rate, which equals the throughput. 
\begin{lemma}\label{lemma:1D-flow}
    For the one-dimensional ISAC network, for any feasible $(s,f)$, we must have $f_{j,j+1} = f, \forall j \in [1:K-1]$.
\end{lemma}
\begin{IEEEproof}
    Applying \eqref{eq:flow} at node $j$, we have $f_{j-1,j} + f_{j+1,j} - f_{j,j-1} - f_{j,j+1} = 0$. Due to Remark~\ref{remark:one-link-off}, either $f_{j+1,j}$ or $f_{j,j+1}$ must be zero. If $f_{j,j-1} = 0$ and $f_{j-1,j} > 0$, then  $f_{j+1,j} = 0$ and $f_{j,j+1} = f_{j-1,j}$.

    By definition, $f_{2,1} = 0$, and thus $f_{1,2} = f$. By induction, for $j =2,3,\dotsc, K-1$, we complete the proof.
\end{IEEEproof}

From Remark~\ref{remark:one-link-off}, the ``backward'' links are not allocated any capacity and carry zero communication rate.

Define $c_\text{min} := \min_{u \in U} c_u$. We now derive the maximum feasible throughput:
\begin{lemma}\label{lemma:1D-max-througput}
    For the one-dimensional ISAC network, $f^* = c_\text{min}$.
\end{lemma}
\begin{IEEEproof}
    Note that $c_\text{min} = \min_{j \in [1:K-1]} c_{\{j,j+1\}}$. \underline{Achievability:}
    From Lemma~\ref{lemma:max-throughput}, we know that $f^*$ can be achieved at $s=0$. We set $s_e = 0, \forall e \in E$. Next, we set $f_{j,j+1} = c_\text{min}$ and $f_{j+1,i} = 0$ for all $j \in [1:K-1]$. This set of rate assignments satisfy \eqref{eq:link-capacity-splitting}, \eqref{eq:trade-off}, and \eqref{eq:flow}, where the link split is $c_{j,j+1} = c_{\{j,j+1\}}$ and $c_{j+1,j} = 0$ for all $j \in [1:K-1]$. So, the rate assignments are valid, and $(0,c_\text{min})$ is feasible.

    \underline{Converse:} Suppose that $(s',f')$ is feasible for some $f' > c_\text{min}$. From Lemma~\ref{lemma:1D-flow}, the communication-rate assignments must be $f_{j,j+1} = f', \forall j \in [1:K-1]$. They must violate \eqref{eq:link-capacity-splitting} and \eqref{eq:trade-off} for at least one link $\{j,j+1\} \in \arg\!\min_{u \in U} c_u$. This leads to a contradiction, as the rate assignments are not valid.
\end{IEEEproof}

Next, we derive the maximum sensing fidelity for any given feasible throughput. The result then traces the boundary point of the sensing--throughput region $R$.
\begin{lemma}\label{lemma:1D-optimal-trade-off}
    Consider a one-dimentional ISAC network. For any fixed $f \in [0,f^*]$,
    \begin{equation}
        \max \{s: (s,f) \in R\} = \sum_{u \in U(A)} c_u - |U(A)|f,
    \end{equation}
    which is achieved by setting $s_{j,j+1} = c_{\{j,j+1\}} - f$ and $s_{j+1,j} = 0$ for all links $\{j,j+1\} \in U(A)$ in the sensing region.
\end{lemma}

\begin{IEEEproof}
    For $f=0$, the result follows directly from Lemmas~\ref{lemma:max-sensing} and \ref{lemma:max-sensing-value}. Consider $f>0$. We show the following for all $j \in [1:K-1]$. From Lemma~\ref{lemma:1D-flow}, we know that $f_{j,j+1}=f$. From Remark~\ref{remark:one-link-off}, we can set $f_{j+1,j} = s_{j+1,j} = c_{j+1,j} = 0$. Constraints \eqref{eq:link-capacity-splitting} and \eqref{eq:trade-off} dictate that $s_{j,j+1} \leq c_{\{j,j+i\}} - f$. Since $s = \sum_{\{i,j\} \in U(A)} \left( s_{i,j} + s_{j,i} \right)$, where the sensing rate for the ``backward'' link is zero, $s$ must be maximized at $\sum_{u \in U(A)}  \left( c_u - f \right) = \sum_{u \in U(A)} c_u - |U(A)|f$.
\end{IEEEproof}

\begin{figure}[t]
    \centering
    \includegraphics[scale=0.6]{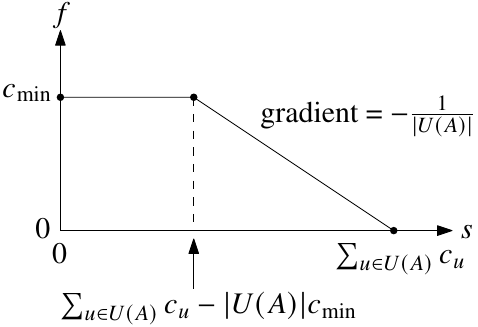}
    \caption{The sensing-throughput region of the  one-dimensional ISAC network}
    \label{fig:1D-region}
\end{figure}

Finally, the sensing--throughput region $R$ is given as follows:
\begin{theorem}\label{theorem:1D-region}
    Consider the one-dimensional ISAC network parameterized by $K$ and $\{c_{\{j,j+1\}}: j \in [1:K-1]\}$. The sensing-throughput region $R$ is given in Figure~\ref{fig:1D-region}.
\end{theorem}

\begin{IEEEproof}
    The result follows directly from Lemmas~\ref{lemma:1D-max-througput} and \ref{lemma:1D-optimal-trade-off}.
\end{IEEEproof}

\subsection{Discussion}

\subsubsection{Free Sensing}
We define free sensing as any $s > 0$ for which $(s,f^*) \in R$, i.e., the amount of non-zero sensing fidelity possible even when the throughput is already maximized.

\begin{lemma}
    In the one-dimensional network, free sensing occurs if and only if $\exists u \in U(A)$ s.t. $c_u > c_\text{min}$.
\end{lemma}

\begin{IEEEproof}
    From Theorem~\ref{theorem:1D-region}, free sensing occurs when $\sum_{u \in U(A)} c_u > |U(A)|c_\text{min}$. Since $c_u \geq c_\text{min}, \forall u \in U$ by definition, the inequality for free sensing holds if and only if $c_u > c_\text{min}$ for some $u \in U(A)$.
\end{IEEEproof}

\subsubsection{Free Communication}
Similarly, we define free communication as any $f > 0$ for which $(s^*,f) \in R$, i.e., the amount of non-zero throughput possible even when the sensing fidelity is already maximized.

We see from Theorem~\ref{theorem:1D-region} that no free communication is possible in the one-dimensional network. The reason is that any non-zero communication must incur a strictly positive communication rate in all forward links, which thus ``uses'' up some amount of all link capacities. Consequently, the links in $U(A)$ cannot be used solely for sensing, which is required to achieve maximum sensing fidelity.

\subsubsection{Optimal Sensing--Throughput Trade-Off}
From Theorem~\ref{theorem:1D-region}, the optimal trade-off between the sensing fidelity and the throughput (i.e., on the Pareto boundary) is $\frac{1}{|U(A)|}$, where $|U(A)|$ is the ``size'' of the sensing region measured by the number of undirected links therein.

The reason for this trade-off is that any increase of $\delta$ amount of communication exhausts every undirected link capacity by $\delta$. As all remaining link capacities are used for sensing (to maximize sensing for the given throughput rate), the total amount of sensing in the sensing region must then reduce by $\delta$ times the number of undirected links in the sensing region.

\subsubsection{An Example}
\begin{figure}[t]
    \centering
    \includegraphics[scale=0.7]{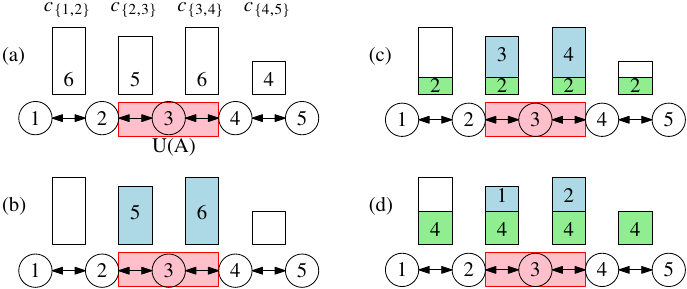}
    \caption{Subfigure (a) shows a one-dimensional ISAC network example. Subfigures (b)--(d) shows different operating points, where the green boxes represent communication rates, and blue boxes, sensing rates.}
    \label{fig:1D-example-rates}
\end{figure}

\begin{figure}[t]
    \centering
    \includegraphics[scale=0.6]{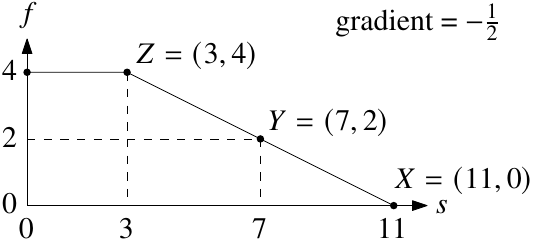}
    \caption{The sensing--throughput region of the network in Figure~\ref{fig:1D-example-rates}}
    \label{fig:1D-example-region}
\end{figure}
Consider a one-dimensional ISAC network with $K=5$ nodes and $U(A) = \{ \{2,3\}, \{3,4\} \}$, where $c_{\{1,2\}}=6$, $c_{\{2,3\}}=5$, $c_{\{3,4\}}=6$, $c_{\{4,5\}}=4$, as depicted in Figure~\ref{fig:1D-example-rates}(a). Figure~\ref{fig:1D-example-region} shows the sensing-throughput region of the network. In particular, feasible points $X$, $Y$, and $Z$ are achieved using rate assignments shown in Figures~\ref{fig:1D-example-rates}(b), (c), and (d), respectively. 

\section{The General ISAC Network}

For the general ISAC network, an analytical expression for the sensing--throughput region remains elusive. However, we will derive several important results to gain an understanding of the Pareto boundary.

\subsection{Free Communication}

Although free communication is not possible in the one-dimensional network, it may be possible in the general network. Define the maximum free communication throughput as $\tilde{f} := \max\{ f: (s^*, f) \in R \}$.

First of all, we know that the point $(s^*, 0) \in R$, where $s^* =\sum_{u \in U(A) c_u}$. The sensing fidelity $s^*$ is achieved if and only if all the links in $U(A)$ are for sensing, i.e., operating at zero-communication rate on them:
\begin{equation}\label{eq:zero-comm-rates}
    f_e = 0,  \forall e \in E(A).
\end{equation}

With this condition in place, the maximum throughput can be achieved by using all other links solely for communications. The maximum throughput can then be achieved by solving $\mathcal{P}_1(G,\{c_u:u \in U\}, A, s^*)$, or equivalently a slightly simpler linear program 
$\mathcal{P}_2(G,\{c_u:u \in U\}, A)$:
\begin{subequations}\label{eq:linear-program-2}
\begin{align}
 \max_{f_e: e \in E \setminus E(A)} \, \, & f\\
\textrm{s.t.} \, \, & \eqref{eq:flow}, \eqref{eq:zero-comm-rates},\\
& f_e \geq 0,  \forall e \in E \setminus E(A),\\
& f_{\{i,j\}} + f_{\{j,i\}} \leq c_{\{i,j\}},  \forall \{i,j\} \in U \setminus U(A).
\end{align}
\end{subequations}

The above observation gives the following results of free communication:
\begin{lemma}\label{lemma:general-throughput-for-max-sensing}
    Consider a general ISAC network. Let $P$ be the optimal value for the linear program $\mathcal{P}_1(G,\{c_u:u \in U\}, A, s^*)$, or equivalently, 
$\mathcal{P}_2(G,\{c_u:u \in U\}, A)$. We have the following:
    \begin{enumerate}
        \item $\tilde{f} = P$.
        \item The straight line joining  $(0,s^*)$ and $(P,s^*)$ are on the Pareto boundary of $R$.
        \item $\tilde{f} > 0$ if and only if there exists a path from the source to the destination that does not intersect with $E(A)$.
    \end{enumerate}
\end{lemma}

\begin{IEEEproof}
    Points 1) and 2) follow directly from the observations preceding the lemma. For Point 3), we first note that all links in $E(A)$ have been fully used by sensing and hence zero communication rate. If a non-zero throughput can be achieved, then it must be supported by communication rates along a path from the source to the destination outside $E(A)$. Conversely, if such a path exists, then a non-zero communication rate can be assigned to this path to achieve a non-zero throughput. 
\end{IEEEproof}

\begin{remark}
    Point~3) in Lemma~\ref{lemma:general-throughput-for-max-sensing} also proves that free communication is impossible in the one-dimensional network if $A \neq \emptyset$.
\end{remark}

\subsection{Free Sensing}

The maximum free sensing fidelity $\tilde{s} := \max\{ (s: (s,f^*) \in R \}$ can be found by iterating the linear program $\mathcal{P}_1$. Define $v(s')$ to be the optimal solution of $\mathcal{P}_1(G,\{c_u:u \in U\}, A, s')$. From Lemmas~\ref{lemma:max-throughput} and \ref{lemma:lp}, we can find $f^* = v(0)$.

\SetKwComment{Comment}{/* }{ */}
\RestyleAlgo{ruled}
\begin{algorithm}
\caption{An algorithm to approximate $\tilde{s}$}\label{algo:bisection}
\SetKwInOut{Input}{input}\SetKwInOut{Output}{output}
\Input{$G, \{c_u: u \in U\}, A, \Delta >0$, $f^* = v(0)$}
\Output{s}

$L \gets 0$\;
$U \gets s^*$\;
$M \gets 0$\;
\While{$U-L>\Delta$}{
    $M\leftarrow (L+U)/2$\;
    Solve $\mathcal{P}_1(G,\{c_u:u \in U\}, A, M)$ and get $v(M)$\;
  \eIf{$v(M) = f^*$}{
    $L\leftarrow M$\;
  }{
      $U\leftarrow M$\;
  }
}
$s \gets L$\;
\end{algorithm}

\begin{lemma}\label{lemma:approx-max-free-sensing}
    Consider a general ISAC network. Algorithm~\ref{algo:bisection} finds $s'$, where $s' \geq \tilde{s} - \Delta$, for some $\Delta > 0$ after running the linear program $\mathcal{P}_1$ at most $\left(\log_2\left(\frac{s^*}{\Delta}\right)+1\right)$ times.
\end{lemma}

\begin{IEEEproof}
    From Lemma~\ref{lemma:R-convex}, we know that $v(s') = \max\{ f: (f,s') \in R\}$ is non-increasing in $s'$. Algorithm~\ref{algo:bisection} applies a bisection search over $s' \in [0,s^*]$ to identify the largest value $s'$ such that $v(s') = v(0) = f^*$ within the accuracy of $\Delta$. This proves $s' \geq \tilde{s} - \Delta$. 

    After $k$ iterations of the algorithm, the interval $L-U$ is $\frac{s^*}{2^k}$. Since the algorithm terminates when $U-L = \frac{s^*}{2^k} \leq \Delta$, any $k \geq \log_2(\frac{s^*}{\Delta})$ is sufficient.  So, $\lceil \log_2(\frac{s^*}{\Delta}) \rceil$ iterations plus the initial run to get $v(0)$ are sufficient.
\end{IEEEproof}

\begin{remark}\label{remark:free-sensing}
    Non-zero free sensing, $\tilde{s} > 0$ can be obtained if and only if $f^*$ can be achieved without exhausting all the link capacities in the sensing region. That is, there exists a set of communication-rate assignments $\{f_e:e\in E\}$ that achieve $f^*$, and $f_{i,j} + f_{j,i} < c_{\{i,j\}}$ for some $\{i,j\} \in U(A)$.
\end{remark}

\subsection{Sensing--Throughput Trade-Off}

From Lemmas~\ref{lemma:general-throughput-for-max-sensing} and \ref{lemma:approx-max-free-sensing}, we obtain two points in the sensing--throughput region. They are points $X$ and $Z$ in Figure~\ref{fig:2D-structure}, respectively.

Now, we proceed to find the Pareto boundary between points $X$ and $Z$. At point~$X$, the links outside the sensing region $U(A)$ have been optimized for communications, and all the links within $U(A)$ have been used for sensing. To increase an infinitesimal amount $\delta$ in throughput requires additional communication paths through the region $U(A)$. As we traverse the Pareto boundary from $X$ to $Z$, each increase in throughput will decrease the sensing fidelity in the smallest amount possible. Let $k_1 \in \mathbb{Z}^+$ be the smallest number of intersections between the additional communication paths with $U(A)$. Since $k_1$ links in $U(A)$ must be changed from sensing to communications, the sensing fidelity must reduce by $k_1 \delta$. This gives the line from $X$ to $Y_1$ a gradient of $-\frac{1}{k_1}$.

When we reach $Y_1$, any further infinitesimal increment in throughput cannot be done at the expense of using paths that only intersect $k_1$ links in $U(A)$ anymore. Let $k_2 \in \mathbb{Z}^+$ be the next smallest number of intersections between the required additional communication paths with $U(A)$. Using the same argument, we derive the gradient of the line from $Y_1$ to $Y_2$ to be $-\frac{1}{k_2}$.

The argument is repeated until we finally reach point $Z$.

\subsection{Structure of Sensing--Throughput Region}

\begin{figure}[t]
    \centering
    \includegraphics[scale=0.6]{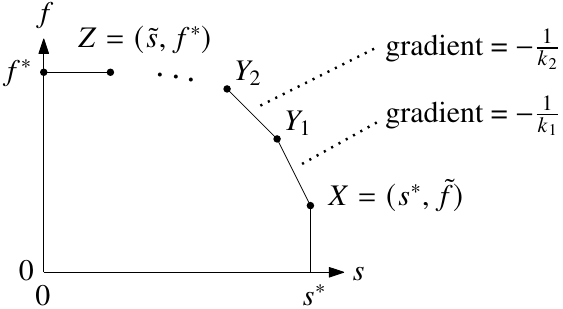}
    \caption{The structure of the sensing--throughput region for the general ISAC network.}
    \label{fig:2D-structure}
\end{figure}

From the above results, we can then derive the structure of the sensing--throughput region for the general ISAC network:
\begin{theorem}
    Consider a general ISAC network. $R$ follows the shape of Figure~\ref{fig:2D-structure}, in which
    \begin{itemize}
        \item Point $X = (s^*, \tilde{f})$, where $\tilde{f}$ is given by Lemma~\ref{lemma:general-throughput-for-max-sensing}.
        \item Point $Z = (\tilde{s}, f^*)$, where $\tilde{s}$ is approximated by Lemma~\ref{lemma:approx-max-free-sensing}.
        \item The lines $(X \text{ to } Y_1), (Y_1 \text{ to } Y_2), \dotsc, (Y_\ell \text{ to } Z)$ are piece-wise linear, where the gradients are $-\frac{1}{k_1}, -\frac{1}{k_2}, \dotsc, -\frac{1}{k_\ell}$, respectively. Here, $k_i \in \mathbb{Z}^+, \forall i \in [1:\ell]$, is the smallest number of links in $U(A)$ that the incremental throughput needs to use.
\end{itemize}
\end{theorem}

\subsection{An Illustrative Example}
\begin{figure}[!t]
    \centering
    \includegraphics[scale=1.0]{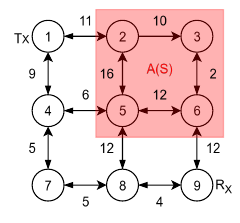}
    \caption{The sensing--throughput region of the network in Figure~\ref{fig:2D-example-rates}.}
    \label{fig:2D-example-rates}
\end{figure}
\begin{figure}[!t]
    \centering
    \includegraphics[scale=0.6]{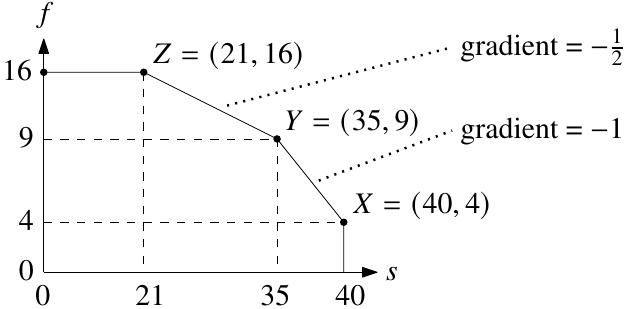}
    \caption{The sensing--throughput region of the network in Figure~\ref{fig:2D-example-rates}.}
    \label{fig:2D-example-region}
\end{figure}

\begin{figure}[!t]
    \centering
    \includegraphics[scale=1.0]{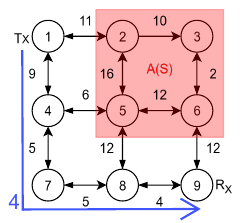}
    \caption{Communication rates to achieve Point $X$ in Figure~\ref{fig:2D-example-region}, where all communication paths avoid the sensing region altogether, giving the maximum sensing fidelity $s^* = 40$.}
    \label{fig:2D-example-rate-1}
\end{figure}

\begin{figure}[!t]
    \centering
    \includegraphics[scale=1.0]{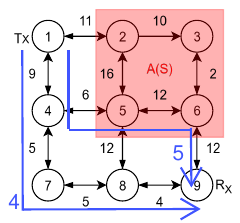}
    \caption{Communication rates to achieve Point $Y$ in Figure~\ref{fig:2D-example-region}, where a communication path interferes with one link in $U(A)$.}
    \label{fig:2D-example-rate-2}
\end{figure}

\begin{figure}[!t]
    \centering
    \includegraphics[scale=1.0]{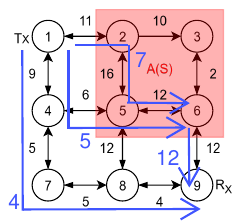}
    \caption{Communication rates to achieve Point $Z$ in Figure~\ref{fig:2D-example-region}. Two communication paths pass through $U(A)$: the lower intersects $U(A)$ in one link, whereas the upper in two links. Here, the throughput is maximized, as all links to $R_x$ are fully used for communications.}
    \label{fig:2D-example-rate-3}
\end{figure}

Consider a general ISAC network depicted in Figure~\ref{fig:2D-example-rates}. The sensing--throughput region is given in Figure~\ref{fig:2D-example-region}, which was numerically generated by solving $\mathcal{P}_1$ over $T_S \in [0:s^*]$. Points $X$, $Y$, and $Z$ can be achieved by using the communication rates in Figures~\ref{fig:2D-example-rate-1}, \ref{fig:2D-example-rate-2}, and \ref{fig:2D-example-rate-3} respectively. 

Along the line from $X$ (exclusive) to $Y$ (inclusive), a communication path intersects with one link in $U(A)$, given the gradient of $-1$. Along the line from $Y$ (exclusive) to $Z$ (inclusive), a communication path intersects with two links in $U(A)$, given the gradient of $-\frac{1}{2}$.

\section{Conclusions}
This paper investigated the fundamental performance limits of network-level ISAC by characterizing the trade-off between communication throughput and sensing coverage in relay-assisted networks. By moving beyond link-level analysis, we formulated a framework that integrates multi-node routing with spatial sensing regions, providing a complete analytical characterization of the sensing–throughput region for one-dimensional paths. For general network topologies, our results demonstrate that the Pareto boundary of this region is piecewise linear, revealing how specific routing strategies and node configurations dictate the optimal balance between dual functions.

\newpage

\balance

\bibliography{sample}

\end{document}